\documentstyle[12pt]{article}

\begin{document}


\begin{center}
Extension of the Ginibre Ensembles of Random Matrices 
\end{center}
\begin{center}
Maciej M. Duras
\end{center}
\begin{center}
Institute of Physics, Cracow University of Technology, 
ulica Podchor\c{a}\.zych 1, PL-30084 Cracow, Poland
\end{center}

\begin{center}
Email: mduras @ riad.usk.pk.edu.pl
\end{center}

\begin{center}
"The 33rd Symposium on Mathematical Physics (SMP33)"; 
June 5th, 2001 - June 9th, 2001; 
Institute of Physics, Nicholas Copernicus University, 
Toru\'{n}, Poland.
\end{center}

\section{Abstract}
\label{sec-Abstract}

The Ginibre ensemble of nonhermitean random Hamiltonian
matrices $K$ is considered.
Each quantum system described by $K$ is a dissipative system 
and the eigenenergies $Z_{i}$
of the Hamiltonian are complex-valued random variables.
The second difference of complex eigenenergies
is viewed as discrete analog of Hessian with respect to labelling index.
The results are considered in view of Wigner and Dyson's
electrostatic analogy.
An extension of space of dynamics of random magnitudes
is performed by introduction of discrete space of labeling
indices.
The comparison with the Gaussian ensembles of random
hermitean Hamiltonian matrices $H$ is performed. 

\section{Introduction}
\label{sec-introduction}

Random Matrix Theory RMT studies
quantum Hamiltonian operators $H$
which are random matrix variables.
Their matrix elements
$H_{ij}$ are independent random scalar variables
\cite{Haake 1990,Guhr 1998,Mehta 1990 0,Reichl 1992,Bohigas 1991,Porter 1965,Brody 1981,Beenakker 1997}.
There were studied among others the following
Gaussian Random Matrix ensembles GRME:
orthogonal GOE, unitary GUE, symplectic GSE,
as well as circular ensembles: orthogonal COE,
unitary CUE, and symplectic CSE.
The choice of ensemble is based on quantum symmetries
ascribed to the Hamiltonian $H$. The Hamiltonian $H$
acts on quantum space $V$ of eigenfunctions.
It is assumed that $V$ is $N$-dimensional Hilbert space
$V={\bf F}^{N}$, where the real, complex, or quaternion
field ${\bf F}={\bf R, C, H}$,
corresponds to GOE, GUE, or GSE, respectively.
If the Hamiltonian matrix $H$ is hermitean $H=H^{\dag}$,
then the probability density function of $H$ reads:
\begin{eqnarray}
& & f_{{\cal H}}(H)={\cal C}_{H \beta} 
\exp{[-\beta \cdot \frac{1}{2} \cdot {\rm Tr} (H^{2})]},
\label{pdf-GOE-GUE-GSE} \\
& & {\cal C}_{H \beta}=(\frac{\beta}{2 \pi})^{{\cal N}_{H \beta}/2}, 
\nonumber \\
& & {\cal N}_{H \beta}=N+\frac{1}{2}N(N-1)\beta, \nonumber \\
& & \int f_{{\cal H}}(H) dH=1,
\nonumber \\
& & dH=\prod_{i=1}^{N} \prod_{j \geq i}^{N} 
\prod_{\gamma=0}^{D-1} dH_{ij}^{(\gamma)}, \nonumber \\
& & H_{ij}=(H_{ij}^{(0)}, ..., H_{ij}^{(D-1)}) \in {\bf F}, \nonumber
\end{eqnarray}
where the parameter $\beta$ assume values 
$\beta=1,2,4,$  for GOE($N$), GUE($N$), GSE($N$), respectively,
and ${\cal N}_{H \beta}$ is number of independent matrix elements
of hermitean Hamiltonian $H$.
The Hamiltonian $H$ belongs to Lie group of hermitean $N \times N {\bf F}$-matrices,
and the matrix Haar's measure $dH$ is invariant under
transformations from the unitary group U($N$, {\bf F}).
The eigenenergies $E_{i}, i=1, ..., N$, of $H$, are real-valued
random variables $E_{i}=E_{i}^{\star}$.
It was Eugene Wigner who firstly dealt with eigenenergy level repulsion
phenomenon studying nuclear spectra \cite{Haake 1990,Guhr 1998,Mehta 1990 0}.
RMT is applicable now in many branches of physics:
nuclear physics (slow neutron resonances, highly excited complex nuclei),
condensed phase physics (fine metallic particles,  
random Ising model [spin glasses]),
quantum chaos (quantum billiards, quantum dots), 
disordered mesoscopic systems (transport phenomena),
quantum chromodynamics, quantum gravity, field theory.

\section{The Ginibre ensembles}
\label{sec-ginibre-ensembles}

Jean Ginibre considered another example of GRME
dropping the assumption of hermiticity of Hamiltonians
thus defining generic ${\bf F}$-valued Hamiltonian $K$
\cite{Haake 1990,Guhr 1998,Ginibre 1965,Mehta 1990 1}.
Hence, $K$ belong to general linear Lie group GL($N$, {\bf F}),
and the matrix Haar's measure $dK$ is invariant under
transformations form that group.
The distribution of $K$ is given by:
\begin{eqnarray}
& & f_{{\cal K}}(K)={\cal C}_{K \beta} 
\exp{[-\beta \cdot \frac{1}{2} \cdot {\rm Tr} (K^{\dag}K)]},
\label{pdf-Ginibre} \\
& & {\cal C}_{K \beta}=(\frac{\beta}{2 \pi})^{{\cal N}_{K \beta}/2}, 
\nonumber \\
& & {\cal N}_{K \beta}=N^{2}\beta, \nonumber \\
& & \int f_{{\cal K}}(K) dK=1,
\nonumber \\
& & dK=\prod_{i=1}^{N} \prod_{j=1}^{N} 
\prod_{\gamma=0}^{D-1} dK_{ij}^{(\gamma)}, \nonumber \\
& & K_{ij}=(K_{ij}^{(0)}, ..., K_{ij}^{(D-1)}) \in {\bf F}, \nonumber
\end{eqnarray}
where $\beta=1,2,4$, stands for real, complex, and quaternion
Ginibre ensembles, respectively.
Therefore, the eigenenergies $Z_{i}$ of quantum system 
ascribed to Ginibre ensemble are complex-valued random variables.
The eigenenergies $Z_{i}, i=1, ..., N$,
of nonhermitean Hamiltonian $K$ are not real-valued random variables
$Z_{i} \neq Z_{i}^{\star}$.
Jean Ginibre postulated the following
joint probability density function 
of random vector of complex eigenvalues $Z_{1}, ..., Z_{N}$
for $N \times N$ Hamiltonian matrices $K$ for $\beta=2$
\cite{Haake 1990,Guhr 1998,Ginibre 1965,Mehta 1990 1}:
\begin{eqnarray}
& & P(z_{1}, ..., z_{N})=
\label{Ginibre-joint-pdf-eigenvalues} \\
& & =\prod _{j=1}^{N} \frac{1}{\pi \cdot j!} \cdot
\prod _{i<j}^{N} \vert z_{i} - z_{j} \vert^{2} \cdot
\exp (- \sum _{j=1}^{N} \vert z_{j}\vert^{2}),
\nonumber
\end{eqnarray}
where $z_{i}$ are complex-valued sample points
($z_{i} \in {\bf C}$).
 
We emphasize here Wigner and Dyson's electrostatic analogy.
A Coulomb gas of $N$ unit charges moving on complex plane (Gauss's plane)
{\bf C} is considered. The vectors of positions
of charges are $z_{i}$ and potential energy of the system is:
\begin{equation}
U(z_{1}, ...,z_{N})=
- \sum_{i<j} \ln \vert z_{i} - z_{j} \vert
+ \frac{1}{2} \sum_{i} \vert z_{i}^{2} \vert. 
\label{Coulomb-potential-energy}
\end{equation}
If gas is in thermodynamical equilibrium at temperature
$T= \frac{1}{2 k_{B}}$ 
($\beta= \frac{1}{k_{B}T}=2$, $k_{B}$ is Boltzmann's constant),
then probability density function of vectors of positions is 
$P(z_{1}, ..., z_{N})$ Eq. (\ref{Ginibre-joint-pdf-eigenvalues}).
Therefore, complex eigenenergies $Z_{i}$ of quantum system 
are analogous to vectors of positions of charges of Coulomb gas.
Moreover, complex-valued spacings $\Delta^{1} Z_{i}$
of complex eigenenergies of quantum system:
\begin{equation}
\Delta^{1} Z_{i}=Z_{i+1}-Z_{i}, i=1, ..., (N-1),
\label{first-diff-def}
\end{equation}
are analogous to vectors of relative positions of electric charges.
Finally, complex-valued
second differences $\Delta^{2} Z_{i}$ of complex eigenenergies:
\begin{equation}
\Delta ^{2} Z_{i}=Z_{i+2} - 2Z_{i+1} + Z_{i}, i=1, ..., (N-2),
\label{Ginibre-second-difference-def}
\end{equation}
are analogous to
vectors of relative positions of vectors
of relative positions of electric charges.

The eigenenergies $Z_{i}=Z(i)$ can be treated as values of function $Z$
of discrete parameter $i=1, ..., N$.
The "Jacobian" of $Z_{i}$ reads:
\begin{equation}
{\rm Jac} Z_{i}= \frac{\partial Z_{i}}{\partial i}
\simeq \frac{\Delta^{1} Z_{i}}{\Delta^{1} i}=\Delta^{1} Z_{i}.
\label{jacobian-Z}
\end{equation}
We readily have, that the spacing is an discrete analog of Jacobian,
since the indexing parameter $i$ belongs to discrete space
of indices $i \in I=\{1, ..., N \}$. Therefore, the first derivative
with respect to $i$ reduces to the first differential quotient.
The Hessian is a Jacobian applied to Jacobian.
We immediately have the formula for discrete "Hessian" for the eigenenergies $Z_{i}$:
\begin{equation}
{\rm Hess} Z_{i}= \frac{\partial ^{2} Z_{i}}{\partial i^{2}}
\simeq \frac{\Delta^{2} Z_{i}}{\Delta^{1} i^{2}}=\Delta^{2} Z_{i}.
\label{hessian-Z}
\end{equation}
Thus, the second difference of $Z$ is discrete analog of Hessian of $Z$.
One emphasizes that both "Jacobian" and "Hessian"
work on discrete index space $I$ of indices $i$.
The spacing is also a discrete analog of energy slope
whereas the second difference corresponds to
energy curvature with respect to external parameter $\lambda$
describing parametric ``evolution'' of energy levels
\cite{Zakrzewski 1,Zakrzewski 2}.
The finite differences of order higher than two
are discrete analogs of compositions of "Jacobians" with "Hessians" of $Z$.

The eigenenergies $E_{i}, i \in I$, of the hermitean Hamiltonian $H$
are ordered increasingly real-valued random variables.
They are values of discrete function $E_{i}=E(i)$.
The first difference of adjacent eigenenergies is:
\begin{equation}
\Delta^{1} E_{i}=E_{i+1}-E_{i}, i=1, ..., (N-1),
\label{first-diff-def-GRME}
\end{equation}
are analogous to vectors of relative positions of electric charges
of one-dimensional Coulomb gas. It is simply the spacing of two adjacent
energies.
Real-valued
second differences $\Delta^{2} E_{i}$ of eigenenergies:
\begin{equation}
\Delta ^{2} E_{i}=E_{i+2} - 2E_{i+1} + E_{i}, i=1, ..., (N-2),
\label{Ginibre-second-difference-def-GRME}
\end{equation}
are analogous to vectors of relative positions 
of vectors of relative positions of charges of one-dimensional
Coulomb gas.
The $\Delta ^{2} Z_{i}$ have their real parts
${\rm Re} \Delta ^{2} Z_{i}$,
and imaginary parts
${\rm Im} \Delta ^{2} Z_{i}$, 
as well as radii (moduli)
$\vert \Delta ^{2} Z_{i} \vert$,
and main arguments (angles) ${\rm Arg} \Delta ^{2} Z_{i}$.
$\Delta ^{2} Z_{i}$ are extensions of real-valued second differences:
\begin{equation}
\Delta^{2} E_{i}=E_{i+2}-2E_{i+1}+E_{i}, i=1, ..., (N-2),
\label{second-diff-def}
\end{equation}
of adjacent ordered increasingly real-valued eigenenergies $E_{i}$
of Hamiltonian $H$ defined for
GOE, GUE, GSE, and Poisson ensemble PE
(where Poisson ensemble is composed of uncorrelated
randomly distributed eigenenergies)
\cite{Duras 1996 PRE,Duras 1996 thesis,Duras 1999 Phys,Duras 1999 Nap,Duras 2000 JOptB}.
The Jacobian and Hessian operators of energy function $E(i)=E_{i}$
for these ensembles read:
\begin{equation}
{\rm Jac} E_{i}= \frac{\partial E_{i}}{\partial i}
\simeq \frac{\Delta^{1} E_{i}}{\Delta^{1} i}=\Delta^{1} E_{i},
\label{jacobian-E}
\end{equation}
and
\begin{equation}
{\rm Hess} E_{i}= \frac{\partial ^{2} E_{i}}{\partial i^{2}}
\simeq \frac{\Delta^{2} E_{i}}{\Delta^{1} i^{2}}=\Delta^{2} E_{i}.
\label{hessian-E}
\end{equation}
The treatment of first and second differences of eigenenergies
as discrete analogs of Jacobians and Hessians
allows one to consider these eigenenergies as a magnitudes 
with statistical properties studied in discrete space of indices.
The labelling index $i$ of the eigenenergies is
an additional variable of "motion", hence the space of indices $I$
augments the space of dynamics of random magnitudes.


\end{document}